\documentclass[sigplan,10pt]{acmart}
\renewcommand\footnotetextcopyrightpermission[1]{}
\usepackage{xcolor}
\usepackage{graphicx}
\usepackage{array}
\usepackage{float}
\usepackage{etoolbox}
\AtBeginEnvironment{figure}{\setlength{\belowcaptionskip}{-8pt}}
\AtBeginEnvironment{figure*}{\setlength{\belowcaptionskip}{-8pt}}
\newcommand{\logact}{LogAct}
\newcommand{\agentbus}{AgentBus}
\newcommand{\agentbuses}{AgentBuses}
\newcommand{\bus}{bus}
\newcommand{\anonharness}{AnonHarness}
\newcommand{\logclaw}{LogClaw}
\newcommand{\append}{\texttt{append}}
\newcommand{\poll}{\texttt{poll}}
\newcommand{\readnext}{\texttt{read}}
\newcommand{\StInferring}{\textit{Inferring}}
\newcommand{\StVoting}{\textit{Voting}}
\newcommand{\StDeciding}{\textit{Deciding}}
\newcommand{\StExecuting}{\textit{Executing}}
\newcommand{\frontier}{FrontierModel}
\newcommand{\oldermodel}{Target}
\newcommand{\zippydb}{AnonDB}
\newcommand{\agentkernel}{AgentKernel}
\newcommand{\classical}{Classic}
\newcommand{\llmactive}{LLM-Active}
\newcommand{\llmpassive}{LLM-Passive}
\newcommand{\Decider}{Decider}
\newcommand{\Voter}{Voter}
\newcommand{\Driver}{Driver}
\newcommand{\Executor}{Executor}
\usepackage{soul}
\definecolor{promptbg}{gray}{0.92}
\sethlcolor{promptbg}
\newcommand{\prompt}[1]{\hl{#1}}

\title{\logact{}: Enabling Agentic Reliability via Shared Logs}

\author{Mahesh Balakrishnan, Ashwin Bharambe, Davide Testuggine, David Geraghty, David Mao\\ Vidhya Venkat, Ilya Mironov, Rithesh Baradi, Gayathri Aiyer, Victoria Dudin\\Meta}
\affiliation{\country{}}
\email{}

\begin{abstract}
Agents are LLM-driven components that can mutate environments in powerful, arbitrary ways.
Extracting guarantees for the execution of agents in production environments can be challenging due to asynchrony and failures.
In this paper, we propose a new abstraction called \logact{}, where each agent is a deconstructed state machine playing a shared log.
In \logact{}, agentic actions are visible in the shared log before they are executed; can be stopped prior to execution by pluggable, decoupled voters; and recovered consistently in the case of agent or environment failure.
\logact{} enables agentic introspection, allowing the agent to analyze its own execution history using LLM inference, which in turn enables semantic variants of recovery, health check, and optimization.
In our evaluation, \logact{} agents recover efficiently and correctly from failures; debug their own performance; optimize token usage in swarms; and stop all unwanted actions for a target model on a representative benchmark with just a 3\% drop in benign utility.
\end{abstract}

\begin{document}

\maketitle
\pagestyle{plain}
\section{Introduction}

Agentic harnesses are a powerful new class of systems that execute in a tight loop of observation, inference, and action~\cite{yao2023react}.
In each iteration, the agent observes its environment, invokes an LLM inference call to determine the next action, and executes that action on its environment.
The actions taken by an agentic harness can range from static tool calls on external services to dynamically generated, arbitrary code that runs within an interpreter (e.g., as in CodeAct~\cite{wang2024codeact}).

\begin{figure}[t]
\centering
\includegraphics[width=\columnwidth,page=1]{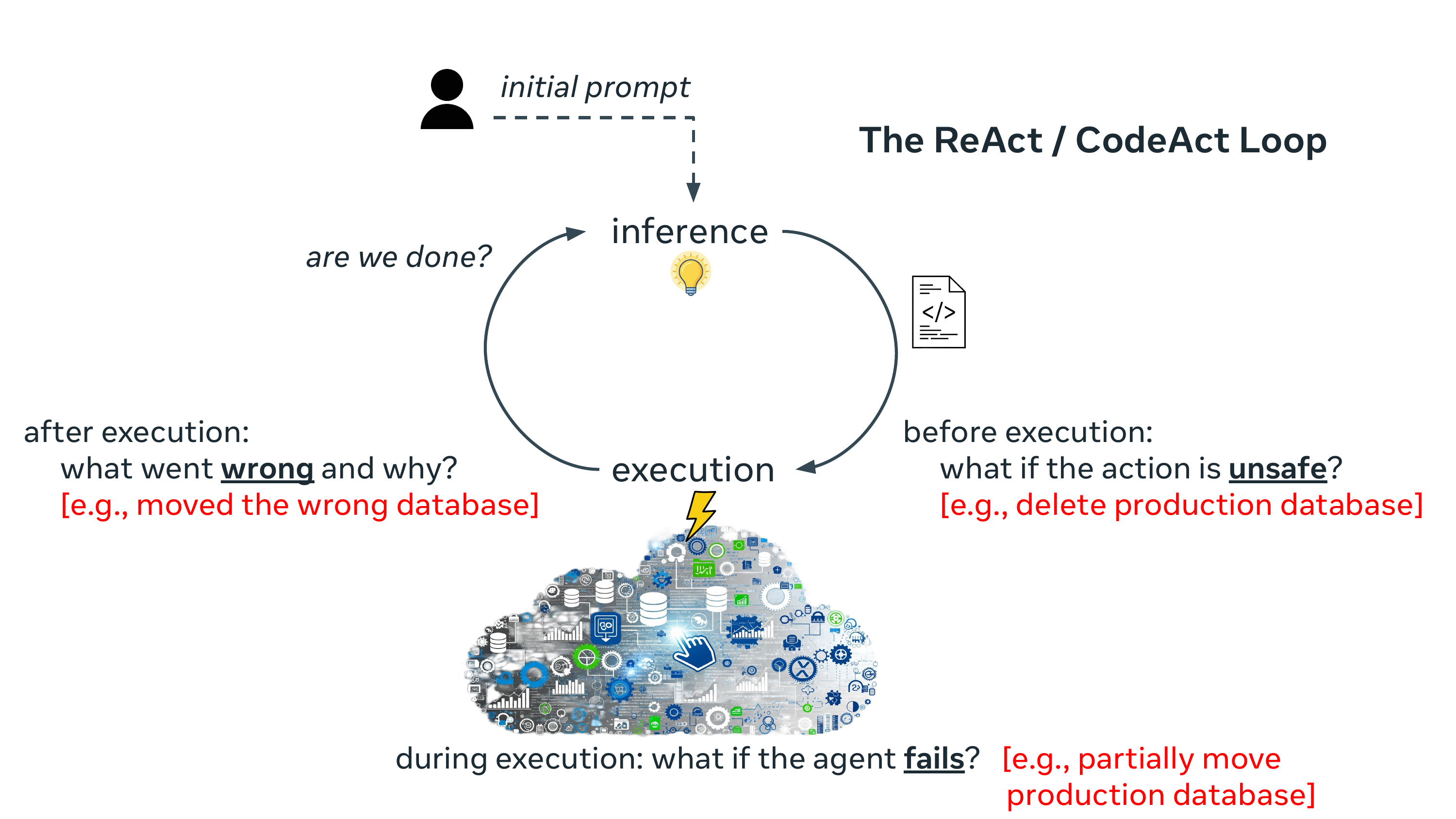}\\[0.5em]
\rule{\columnwidth}{1.5pt}\\[0.5em]
\includegraphics[width=\columnwidth,page=2]{pics/problem.pdf}
\caption{Existing ReAct / CodeAct agents run in an imperative loop; \logact{} is a state machine playing a shared log.}
\label{fig:problem}
\vspace{4pt}
\end{figure}

Unfortunately, extracting strong guarantees from agents is challenging.
In particular, the expressive power of agents makes it difficult to prevent them from taking destructive actions (e.g., deleting critical data).
During execution, failures of the agent or its environment can leave the overall system in an ill-defined state where invariants are broken (e.g., a half-migrated database).
After execution, it can be difficult to determine what an agent did or why.
These concerns are tractable when a single agent operates on a transactional, sandboxed environment (e.g., a coding agent operating on scoped tasks against a git repository); but daunting when agents execute against a shared, distributed environment consisting of many stateful services.
For example, an agent that mistakenly starts deleting a production K8s~\cite{burns2016kubernetes} job and its associated state (telemetry, metadata, databases, etc.) cannot simply stop and restore the entire world to a prior checkpoint.
Importantly, these challenges cannot be overcome just by using more intelligent models or more capable agentic harnesses: even an omniscient and omnipotent agent is subject to failures and asynchrony.

In this paper, we propose the novel abstraction of a \textit{\logact{}} agent: an agent implemented as a state machine over a shared log~\cite{balakrishnan2012corfu, balakrishnan2013tango, lockerman2018fuzzylog, balakrishnan2020delos, balakrishnan2021delos, jia2021boki, luo2024lazylog, bhat2025fixante, zhu2025impeller}.
In \logact{}, \textit{the log is the (logical) agent}: physical machines simply materialize the log by executing \textit{intentions} (i.e., intended commands) stored in it.
An intention must be appended durably to the log -- along with a quorum of votes -- before the agent executes it.
The state machine is \textit{deconstructed} across multiple processes, allowing agentic logic to be implemented as a collection of decoupled components that are isolated from each other.
In turn, the shared log (which we call the \textit{\agentbus{}}) is a remote, durable, append-only log dedicated to each \logact{} agent and shared by these processes, which can concurrently append to it and play entries from it.

In \logact{}, the shared log enables durability and failure atomicity: if an agentic node or its environment experiences a failure, it can repair state by inspecting the log.
The shared log also enables highly available failover: a standby node can take over if a primary node fails.
Finally, the shared log provides an audit trail for all agentic activity: in addition to intentions and votes, it logs inference and action results; mailbox messages from human users and external logical agents; and policy changes for agentic and voter behavior.
In effect, \logact{} combines a number of classical solutions -- shared logs~\cite{balakrishnan2012corfu, balakrishnan2013tango}, State Machine Replication (SMR)~\cite{schneider1990smr}, Write-Ahead Logs (WALs)~\cite{mohan1992aries}, durable workflows~\cite{temporal, restate}, and publish-subscribe~\cite{eugster2003pubsub} -- to address the new problem of agentic fault-tolerance and safety.

If an agent were simply a conventional stateful service, a shared log design would be sufficient to provide safety, fault-tolerance, and audit.
However, agents pose unique challenges for safety and fault-tolerance.
Actions are complex and arbitrary (in the limit, literally lambdas with side-effects), without clear undo/redo schema; and the state they modify spans both the agentic process and its environment.
Failures can be semantically complex, involving not just crashes or unresponsiveness to pings, but also hallucinations, slow or buggy generated code, and complex actions whose safety depends deeply on context, control flow, and history.

To overcome these challenges, we leverage \textbf{\textit{agentic introspection}}: \textit{each \logact{} agent can inspect and run inference -- in real time -- on the past history and imminent intentions stored in its own \agentbus{}}.
Introspection provides a novel solution to each of our challenges.
First, an agent using the \agentbus{} as a WAL to recover from a crash can execute \textit{semantic recovery}: i.e., determine the lambda it was executing at the time of the crash; generate and execute exploratory lambdas to determine the state of the surrounding environment (and consequently the point at which the original lambda was interrupted); and issue compensating actions to either roll forward or roll back.
Second, the physical components of a single logical agent can leverage introspection for safety and availability: for example, an LLM-based safety voter can inspect past activity before stamping an intention; while a warm standby node can perform a \textit{semantic health check} on a primary node before taking over if necessary.

In this paper, we primarily focus on single-agent safety and fault-tolerance; however, \logact{} also provides a foundation for multi-agent coordination.
Each logical agent has its own \agentbus{} instance.
Multiple agents operating on a shared environment can coordinate via mailbox messages on each other's \agentbuses{}, enabling reliable, decoupled communication.
Introspection adds new multi-agent capabilities which can lead to better performance or higher efficiency: agents can inspect each other's logs to build semantic summaries of ongoing activity, detect redundant or conflicting work, and dynamically re-partition tasks across a swarm.

\logact{} has its limitations.
Inference (and hence, introspection) can be arbitrary and difficult to test due to the large input / output space.
As a result, it is difficult to offer a binary guarantee that semantic recovery will restore the system to a clean state; or that an LLM-based voter will always catch an unsafe action.
What \logact{} provides is an architecture to ensure that every agentic action is durable and visible (via the \agentbus{}) to pluggable safety and recovery components.
These components can ensure safety with classical techniques (such as rule-based verifiers, simulators, static analysis, etc.), in which case they offer hard guarantees for well-defined properties.
Or they can be LLM-based components that cover properties more difficult to specify formally or check efficiently.
\logact{} mixes both types of components via quorums, allowing new implementations to be plugged in without downtime, which in turn enables incremental improvements to safety and reliability.

\logact{} is the first proposal for structuring an agent as a state machine playing a shared log; existing systems in this emerging space are typically imperative loops~\cite{yao2023react, wang2024codeact}.
A shared log design enables the new idea of agentic introspection, where the entire execution history of the agent is processed via inference (existing notions of introspection stop at token-only trajectories~\cite{shinn2023reflexion, yao2023react}).
In addition, this paper extends the distributed systems literature in two new ways: adding strong types to shared logs, with access control and selective playback at type grain; and deconstructing a single logical state machine across multiple processes via the shared log.

We make the following contributions in this paper.
\begin{itemize}
\item{We propose a novel abstraction for agentic systems called \logact{}, which replaces the imperative agentic loop with a state machine playing a shared log. In \logact{}, the agentic state machine is deconstructed over multiple processes. A key aspect is agentic introspection, where agents can inspect their own execution for better safety, fault-tolerance, performance, and efficiency.}
\item{We describe different implementations of \logact{}, including clean-slate implementations as well as dirty-slate integrations (e.g., Claude Code), and different \agentbus{} backends (in-memory, SQLite, and \zippydb{}).}
\item{We show that \logact{} enables \textit{safe} agents: a dual-voter setup with a rule-based voter and an LLM-based override stops all unwanted environment actions in the AgentDojo benchmark on a target model, with just a 3\% reduction in benign utility (leaving a residual ASR of 1.4\% for action-less attacks)}.
\item{We show that \logact{} enables \textit{fault-tolerant} agents: in a representative long-running file checksum task, a slow agent is inspected via a semantic health check; and resumed via semantic recovery without re-doing any work; with a code block that is 290X faster.}
\item{We show that \logact{} enables \textit{cheaper, faster} agent swarms: agentic introspection enables a 6-agent swarm to do 17\% more work and use 41\% fewer tokens.}
\end{itemize}

\section{Background}

At first glance, an agentic harness is like any other service.
Users can submit a prompt to it, triggering the start of a turn.
Within the turn, the agent steps through an inner loop, in which it repeatedly invokes LLM inference to determine the next action; and then executes that action on its environment.
In each iteration of this inner loop, the agent examines the result of the action to determine whether the turn is complete or not.

While agentic harnesses started out with access to a relatively narrow set of actions, they have evolved to support arbitrary code execution~\cite{wang2024codeact}.
In the limit, the LLM emits literal code (e.g., Python code blocks) which is executed in an interpreter.
In this form, an agentic harness can be viewed as a self-writing program, generating itself with the LLM's help as it progresses.

Agents are subject to the litany of failure modes that affect conventional distributed systems.
If the physical machine hosting the agent crashes and reboots, any state resident within the process is lost.
If the machine fails permanently, local filesystem state is lost.
A machine can be partitioned away and appear to have failed; only to reappear later when the partition heals.
A machine might fail in the middle of executing a code block within the inner agentic loop, leaving the system in an ill-defined state.

A wealth of distributed systems abstractions and protocols exist to help systems tolerate such failures.
Can we apply them to agents?

\begin{table}[t]
\centering
\caption{Existing fault-tolerance approaches and their limitations for agents. None addresses agentic safety.}
\label{tab:prior-approaches}
\small
\begin{tabular}{>{\raggedright\arraybackslash}p{1.4cm}>{\raggedright\arraybackslash}p{1.8cm}>{\raggedright\arraybackslash}p{2.0cm}>{\raggedright\arraybackslash}p{2.2cm}}
\hline
\textbf{Approach} & \textbf{Key Idea} & \textbf{Assumes} & \textbf{Why Agents Break It} \\
\hline
WAL~\cite{mohan1992aries} & Log before acting & Structured txns with schema + undo/redo & Actions are arbitrary lambdas; no schema \\
\hline
SMR~\cite{schneider1990smr} & Replicated state machine on a log & Full, isolated state copy; deterministic execution & Acts on external environment; non-deterministic \\
\hline
Checkpoint \newline \cite{chandy1985snapshots} & Periodic snapshots; restore on failure & State is local and checkpointable; deterministic & External environment not checkpointable; non-deterministic \\
\hline
Durable Workflows~\cite{temporal, restate} & Record non-det ops; replay on recovery & Predefined control flow; explicit compensating actions & Open-ended control flow (LLM-driven); no predefined undo \\
\hline
\end{tabular}
\end{table}

\textbf{Write-Ahead Logging (WAL)}~\cite{mohan1992aries}\textbf{:} To provide durability and failure atomicity despite machine reboots, a database synchronously writes descriptions of each transaction (i.e., the set of keys read and written by it) to a log before modifying its in-place state.
If the database fails and recovers, it replays these entries and either rolls forward or rolls back any incomplete transactions.
A challenge with applying this approach to agents is that agentic `transactions' can be arbitrary lambdas, without a clear schema or undo/redo actions.

\textbf{State Machine Replication (SMR)}~\cite{schneider1990smr}\textbf{:} For durability and high availability in the event of permanent crashes and network partitions, respectively, services are written as replicated state machines, with multiple physical machines redundantly executing a total order of commands.
Applying SMR directly to agents is difficult, since each agent does not maintain a full, isolated copy of state; instead, it acts upon an external, durable environment.
As a result, SMR-like recovery where a new node can simply play the total order from the beginning (or more practically, find a snapshot of state and play only the subsequent suffix of the log) is infeasible.
In addition, coarse-grained input-only SMR logging is insufficient for agentic recovery since the agent state machine is not deterministic.

\textbf{Checkpointing}~\cite{chandy1985snapshots}\textbf{:} In some distributed systems (including large-scale AI training~\cite{mohan2021checkfreq}), a common solution for fault-tolerance involves checkpointing state periodically.
On a failure, the latest checkpoint is re-loaded and any subsequent computation is re-executed.
Unfortunately, it is likely not feasible to checkpoint the external, durable environment that an agent operates on.
Even if the amount of state were tractable, this might require stopping the world for other agents accessing the same environment.
Further, checkpoint / restore assumes that computation is either deterministic or that determinism is not necessary; whereas for agents, determinism is both important and difficult to achieve (e.g., due to user input, LLM non-determinism, etc.).

\textbf{Durable Workflows}~\cite{temporal, restate}\textbf{:} Systems such as Temporal and Restate operate over external, durable environments.
They aim for fault-tolerant exactly-once execution by recording the results of non-deterministic operations and replaying them deterministically on recovery.
However, they require the developer to define the workflow's control flow graph upfront, including explicit compensating actions for each step.
Agents violate both assumptions: the control flow is open-ended, with each next step determined at runtime by LLM inference over the entire history; and actions are arbitrary lambdas for which no predefined undo exists.

Each of these approaches contains useful principles for agentic fault-tolerance.
WALs provide the idea of logging before acting; SMR introduces the notion of modeling an agent as a single logical state machine over a shared log; Durable Workflows add exactly-once execution via careful recovery logic.
None of these prior approaches addresses agentic safety.
In SMR systems, voters (i.e., acceptors) typically decide on the ordering of commands (or more precisely, the assignment of a command to an ordering slot), not whether those commands are allowed to be executed; whereas Atomic Commit protocols usually predicate votes on database properties like integrity and isolation.
Further, shared logs have provided a common substrate for these different systems: used as WALs~\cite{bernstein2011hyder}, SMR layers~\cite{balakrishnan2013tango}, and durable functions~\cite{jia2021boki}.
Can we use the shared log in a novel way to enable a new type of system that addresses agentic safety and fault-tolerance?

\section{The \logact{} Abstraction}

\begin{figure}[t]
\centering
\includegraphics[width=\columnwidth]{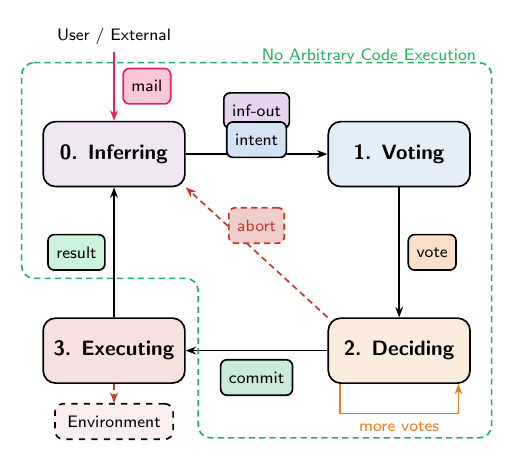}
\caption{The \logact{} state machine. The agent cycles through four stages; each stage is triggered by playing a typed log entry and concludes by appending one.}
\label{fig:state-machine}
\end{figure}

\begin{figure}[t]
\centering
\includegraphics[width=\columnwidth]{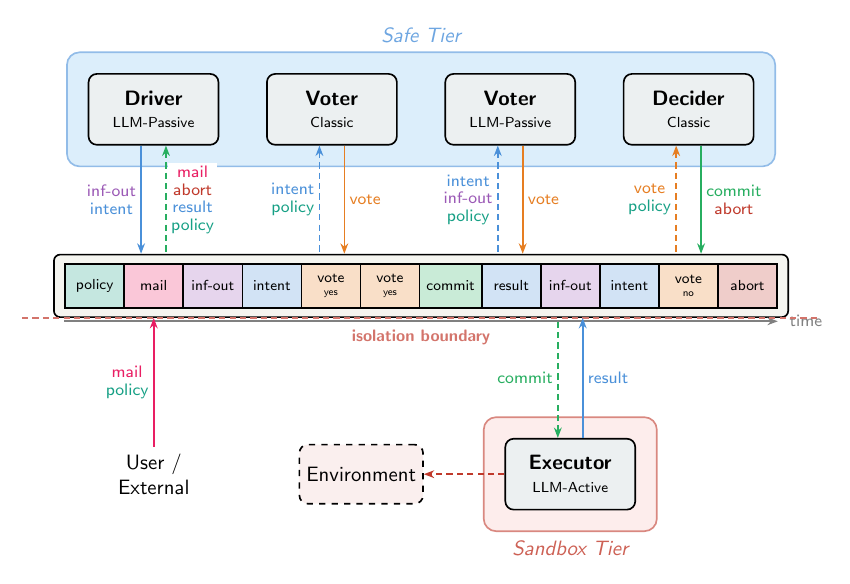}
\caption{A deconstructed \logact{} agent on the \agentbus{}. Physical nodes -- \Driver{}, \Voter{}(s), \Decider{}, and \Executor{} -- each append and play a subset of entry types.}
\label{fig:deconstructed}
\end{figure}

In \logact{}, each logical agent is structured as a state machine playing a shared log.
As the agent alternates between LLM invocations (to determine the next action to take) and actions on its own state and the surrounding environment, it appends each state transition durably to the shared log before executing it.
In this section, we describe the two layers that comprise \logact{}: the \agentbus{}; and a \textit{deconstructed} state machine above it, which distributes the logic of the agent across a set of physical machines appending to / reading from the \agentbus{}.

The \agentbus{} is a linearizable, durable shared log with a standard API~\cite{balakrishnan2020delos}: clients can \append{} entries and obtain the log position at which the entry was written (i.e., its logical timestamp); check the current tail of the log; and read entries within a range of log positions.
The \agentbus{} makes two additions to this basic API.
First, the \agentbus{} introduces a strong notion of types: each entry is tagged with a type (e.g., intention, result, vote, decision, inference output, mailbox message), and the \append{} / \readnext{} calls take optional type parameters.
Second, we add a \textit{blocking poll} API that returns when an entry with a type in a specified set shows up on the \agentbus{}.
Importantly, the API enforces access control at the granularity of types: clients can only \append{}, \readnext{}, or \poll{} the entry types that they have permissions for (e.g., \append{} only votes; \poll{} only intentions).

\begin{figure}[t]
\centering
\small
\begin{verbatim}
struct Entry {
    position:  u64,       // log position
    realtime_ts: u64,     // wall-clock ms
    payload:   Payload,   // typed payload
}

enum PayloadType {
    InfIn, InfOut, Intent, Vote,
    Commit, Abort, Result, Mail, Policy,
}

// Append a typed payload to the log.
fn append(payload: Payload)
    -> u64; // durable log position

// Read entries in [start, end) range.
fn read(start: u64, end: u64)
    -> Vec<Entry>;

// Return the current tail position.
fn tail() -> u64;

// Blocking poll: wait for entries with a
// type in the filter set at position >= start.
fn poll(
    start:  u64,
    filter: Vec<PayloadType>,
) -> Vec<Entry>;
\end{verbatim}
\caption{Pseudocode for the \agentbus{} API. The standard shared log operations are \append{}, \texttt{read}, and \texttt{tail}. The \agentbus{} adds strong types on entries and a blocking \poll{} that waits for entries of specified types.}
\label{fig:api}
\end{figure}

\textbf{The \logact{} agent is implemented as a state machine above the \agentbus{}} (Figure~\ref{fig:state-machine}).
As with any other agentic loop, the \logact{} state machine alternates between sending requests to an inference layer, which responds with proposed actions; verifying those actions with safety checks; and executing those actions on its environment if the checks pass.
However, unlike conventional agentic loops, these state changes are driven entirely by the underlying log:
\begin{enumerate}
\setcounter{enumi}{-1}
\item{\StInferring{}: the agent plays an initial user input or a prior action's result; it invokes the inference layer with this input; it appends an inference output entry along with an extracted intention entry;}
\item{\StVoting{}: the agent plays an intent from the log; verifies its safety; and appends one or more votes;}
\item{\StDeciding{}: the agent plays votes from the log; determines if a quorum of votes has been reached; and appends a commit or an abort;}
\item{\StExecuting{}: the agent plays a commit from the log; executes the corresponding action on its environment; and appends a result to the log, taking it back to the \StInferring{} state.}
\end{enumerate}
If the decision in the \StDeciding{} step is an abort, we skip the \StExecuting{} step and go directly back to \StInferring{}, which also plays aborts from the log.

\textbf{Further, the \logact{} agent is a \textit{deconstructed} state machine above the \agentbus{}} (Figure~\ref{fig:deconstructed}). We split the state machine into multiple components, each of which plays some subset of entry types from the log and appends some other subset of entry types to it.
Concretely, an agentic \textit{\Driver{}} runs the \StInferring{} stage, interacts with inference, and appends intents; \textit{\Voter{}s} run the \StVoting{} stage, playing these intents from the log and appending votes; a \textit{\Decider{}} runs the \StDeciding{} stage, playing these votes and appending commit or abort entries; an \textit{\Executor{}} runs the \StExecuting{} stage, playing commit entries, executing the intention against the environment, and appending the result to the log.
These deconstructed components can be collocated, or isolated on different physical processes or machines.

\textbf{Why deconstruct the state machine?}
We observe that each component has different security, fault-tolerance, and extensibility properties; hence, we benefit by decoupling them and running them separately:
\begin{itemize}
\item{Security: The \Executor{} runs arbitrary LLM-generated code; the other components do not. Separating them allows us to provide strong guarantees (described in Section \ref{sec:guarantees}) even when \Voter{}s make occasional mistakes; while still giving the \Executor{} access to the environment.}
\item{Fault-Tolerance: The \Executor{} state is distributed and difficult to recover; the other components are simple self-contained state machines which support leader election via the log itself (described in Section \ref{sec:failures}).}
\item{Extensibility: \Voter{}s are pluggable / extensible components. We expect multiple \Voter{}s to vote on each intention; the set of \Voter{}s to change over time; and each \Voter{}'s logic and policy to change as well. Decoupling allows us to upgrade \Voter{}s independently from each other and the remaining components.}
\end{itemize}

Two types of control entries originate from outside the agentic state machine:

\textbf{Mailbox:} The \Driver{} plays an additional entry type called mailbox messages (or mail for short), which (among other uses) carries the initial user input.
Any external user (or agent, as part of its own \StExecuting{} stage) can append a mail entry to the log at any time.
If the \Driver{} is quiescent (i.e., not waiting for some prior intention's result from the \Executor{}), any incoming mail immediately triggers an inference call.
Else if the \Driver{} is waiting on the result of some prior intention, any incoming mail is buffered and included in the next inference call.
As a result, external entities can alter the behavior of the logical agent by sending it mail.

\textbf{Policy:} \logact{} configures policy via the log itself. Policy changes are driven via entries on the log, ensuring that they are applied consistently across all the components of the state machine.
\Voter{} policy entries can change the behavior of specific \Voter{}s; e.g., a policy change for an LLM-based \Voter{} might indicate ``allow files in /tmp to be deleted''; or a change for an allowlist \Voter{} might add ``*.tmp'' to the list of files allowed to be deleted.
\Decider{} policy entries change the quorum required for a commit: e.g., we might switch from \texttt{on\_by\_default} (commit without requiring any votes) to \texttt{first\_voter} (decide based on the first voter), or more complex policies that require some specific combination of voter types to agree (\texttt{boolean\_OR} or \texttt{AND} of an LLM-based voter and a rule-based voter).
Policy entries are only allowed from privileged administrative clients, since they can significantly alter the security properties of the system.

\subsection{Safety Guarantees}
\label{sec:guarantees}

We first describe our assumptions; how \logact{} components are isolated; and then the resulting guarantees we obtain.

\textbf{Assumptions:} We assume nothing about the initial user input, which can explicitly request an unsafe action or bias an LLM towards producing one.
Likewise, we make no assumptions about the inference layer: we expect it to emit arbitrary tokens, including unsafe actions as well as prompt injections that can influence downstream LLMs.
We do not trust the target environment either, since it can contain arbitrary state that induces a prompt injection on any LLMs in the system.
Our assumptions are weaker than those of prior work~\cite{debenedetti2025camel}, which assumes that a blank-slate LLM (along with the initial user input) will generate only safe actions unless subject to prompt injection.
In the limit, an LLM can generate arbitrarily unsafe actions and wreak havoc on the environment.

\textbf{Component Isolation:}
Based on their interaction with LLMs (and consequent isolation requirements), we classify \logact{} components as:
\begin{itemize}
\item{\classical{}: No contact with LLMs. E.g., \Voter{}s based on rules, static analysis, simulation, etc.}
\item{\llmpassive{}: Interact with LLMs to send/receive text strings, without executing code. E.g., \Driver{}, LLM-based \Voter{}.}
\item{\llmactive{}: Execute code generated by LLMs. E.g., \Executor{}.}
\end{itemize}

\classical{} and \llmpassive{} components reside on a ``safe'' tier that is inaccessible to \llmactive{} components.
In turn, the \llmactive{} \Executor{} runs in a sandbox tier with selective access to the production environment (to carry out its tasks), but without any access to the safe tier.
\llmpassive{} \Voter{}s are barred from accessing the environment, by default; allowing them access can result in better votes but also compromise safety, as we discuss shortly.
All components can access the \agentbus{}, albeit with access control on the types of entries they can \append{} or \poll{}.

\textbf{Properties:} Let $L = \langle I_1, I_2, \ldots, I_n \rangle$ be the ordered sequence of $n$ committed intentions on the log, and let $E$ be the environment after the \Executor{} supposedly executes $L$. We denote by $L_k = \langle I_1, \ldots, I_k \rangle$ the prefix of $L$ of length $k$, and by $E_k$ the environment after the \Executor{} supposedly executes $L_k$, with $E_0$ being the initial environment. For simplicity, we make an exclusivity assumption: i.e., that $E$ is acted upon solely by the \Executor{} playing $L$; later, we examine the practicality of this assumption. We now define two properties.

\textit{Consistency}: $L$ and $E$ are consistent if for all $k \leq n$, $E_k$ is the result of faithfully executing $L_k$ on $E_0$.

\textit{Safety}: let $\mathcal{S}$ be a set of invariants over the environment, provided by an external user (e.g., a human administrator). $E_k$ is \textit{safe} if it satisfies all invariants in $\mathcal{S}$; $L$ is safe if a faithful execution of $L_k$ results in a safe $E_k$ for all $k \leq n$. We assume $E_0$ satisfies $\mathcal{S}$.

If the \Executor{} faithfully executes each $I_k$, we obtain Consistency: each $E_k$ reflects the faithful execution of $L_k$ on $E_0$. If the \Voter{}s / \Decider{} correctly enforce $\mathcal{S}$ (i.e., only committing intentions whose execution will not violate $\mathcal{S}$), we obtain Safety: any $I_k$ whose execution would cause $E_k$ to violate $\mathcal{S}$ is blocked before reaching the \Executor{}. In practice, however, \Voter{}s can make mistakes; they enforce only an approximation $\tilde{\mathcal{S}}$ of $\mathcal{S}$, yielding a weaker property we call Enforced-Safety.

What happens when $\tilde{\mathcal{S}}$ is not identical to $\mathcal{S}$? An unsafe intention can be approved by the \Voter{} and played by the \Executor{}. There are three possible outcomes:
\begin{itemize}
\item{Case 1: \textit{The \Executor{} modifies the environment, but not itself or the other components.} In this case, the unsafe intention violates Safety, but not Enforced-Safety or Consistency. The log and the environment continue to be consistent with each other, and to provide Enforced-Safety. This is the benign, common case. Note that if $\mathcal{S}$ and $\tilde{\mathcal{S}}$ deviate massively, this can still be disastrous (e.g., delete all of production), but with reasonable \Voter{}s we expect damage to be limited (e.g., modify a single row in some table).}
\item{Case 2: \textit{The \Executor{} modifies itself.} In this case, the unsafe intention can violate Consistency. For example, the \Executor{} runs code that makes it lie about its execution for all future intentions. The environment no longer matches the log, even though the log still satisfies Enforced-Safety.}
\item{Case 3: \textit{The \Executor{} modifies or replaces the \Decider{} / \Voter{}s.} The unsafe intention arbitrarily changes Enforced-Safety, retaining Consistency. This is the worst case: an \Executor{} can swap existing \Voter{}s with its own puppets, change their policies arbitrarily, or rewrite their code, etc. so that all future intentions are passed.}
\end{itemize}

Case 1 is inherently difficult to prevent, unless we have completely infallible \Voter{}s; practically, we expect testing and monitoring to ensure that Enforced-Safety never deviates hugely from Safety.
In addition, we can use quorums of LLM-based \Voter{}s mixed with classical non-LLM \Voter{}s (and \Decider{}s) that are immune to prompt injections.
Another option involves restricting the \Executor{}'s power by running it within a restricted sandbox, etc.; however, this undermines the power of agents to operate in real-world production environments.
For Case 2, since the log continues to satisfy Enforced-Safety, in principle we have the ability to check consistency between the environment and the log (to catch a lying \Executor{}, for instance).

To prevent Case 3, we need to ensure that \Executor{}s cannot access or influence the \Decider{} / \Voter{}s in any way, or impersonate a \Voter{} or \Decider{} by inserting votes or commits on the log.
Both these properties are provided by running the \Executor{} and \Decider{} / \Voter{}s on isolated infrastructure with different identities.
Isolation ensures that a \Voter{} binary cannot be modified by an \Executor{}, for instance; and also that only \Voter{}s can be supplied with permissions that allow them to append votes.
As noted earlier, LLM-based \Voter{}s do not get access to the environment; if such access is required for better votes, it carries the risk of prompt injection if they observe state written by the \Executor{}.

\textbf{Concurrency}:
In practice, enforcing the exclusivity assumption (i.e., that a single \logact{} agent operates exclusively on the environment) requires some form of concurrency control between agents, either sharding the environment across logical agents (e.g., keys A-E in a DB owned by one agent); or assigning roles to agents (e.g., all control DB accesses by one agent); or allowing agents to use locking protocols.
Further, we also need voters to determine that agents have followed such protocols before executing an intention.
In this paper, we focus on the single-agent case, leaving the detailed exploration of multi-agent coordination protocols to future work.
Modeling individual agents as \logact{} state machines gives us a low-level substrate to build such coordination machinery; in much the same way that single-shard consensus protocols~\cite{lamport1998paxos} are a substrate for fault-tolerant, multi-shard atomic commit protocols~\cite{gray2006paxoscommit}.

Even in the single-agent case, the environment is typically updated by external entities (e.g., other services, users, cron jobs) concurrently with the agent. As a result, the environment state when $I_k$ is executed may differ from the state produced by $I_{k-1}$. Let $\hat{E}_k$ denote the actual environment state immediately before $I_k$ is executed. We generalize our definitions accordingly: $L$ and $E$ satisfy Consistency if for all $k$, executing $I_k$ on $\hat{E}_k$ faithfully reflects $I_k$; and $L$ \textit{preserves Safety (or Enforced-Safety)} if for all $k$, whenever $\hat{E}_k$ satisfies $\mathcal{S}$ (or $\tilde{\mathcal{S}}$), executing $I_k$ on $\hat{E}_k$ produces a state that also satisfies $\mathcal{S}$ (or $\tilde{\mathcal{S}}$). These definitions reduce to the original ones when $\hat{E}_k = E_{k-1}$ for all $k$ (i.e., no foreign modifications). Safety and Enforced-Safety become \textit{preservation} properties: external actions may independently violate $\mathcal{S}$ or $\tilde{\mathcal{S}}$, but the agent must never take a safe state to an unsafe one.

Note that with external actions, \Voter{}s cannot base their vote on the environment, since it can change before the intention executes.
For example, a \Voter{} cannot green-light a decrement operation on a register with a non-negative integrity invariant just by observing the current value of the register.
However, \Voter{}s can base their vote on the logic within the intention itself: e.g., whether it correctly locks / unlocks the register and performs a conditional write on it.

\begin{table}[t]
\centering
\caption{Entry types on the \agentbus{}, and which components append and play them.}
\label{tab:entry-types}
\begin{tabular}{@{}l@{\hspace{1.5em}}l@{\hspace{1.5em}}l@{}}
\hline
\textbf{Entry Type} & \textbf{Appended By} & \textbf{Played By} \\
\hline
Mail (inbound) & External entities & \Driver{} \\
Inference output & \Driver{} & \Driver{}, \Voter{}s (opt.) \\
Intention & \Driver{} & \Voter{}s \\
Vote & \Voter{}s & \Decider{}, \Voter{}s (opt.) \\
Commit & \Decider{} & \Executor{} \\
Abort & \Decider{} & \Driver{} \\
Result & \Executor{} & \Driver{} \\
Policy & External entities & All \\
\hline
\end{tabular}
\end{table}

\subsection{Failure and Recovery}
\label{sec:failures}

In the absence of failures, the correctness and liveness of the \logact{} state machine are easy to understand.
However, when a particular component fails (or appears to fail, i.e., not reacting to an incoming entry or appending its own entry type), the state machine can grind to a halt.
In conventional State Machine Replication, the entire state machine is replicated on multiple nodes; each node plays every entry in the shared log and executes it redundantly, mutating its own exclusive copy of the full state of the system.
However, an agent acts upon durable, external state rather than its own local copy; in this case, how do we recover from failures?

One observation is that deconstructing the state machine allows us to react to different types of failures in different ways, depending on where the state of the component lives (internal copy vs. external).
Another simplifying observation is that in a correct \logact{} state machine without failures, we are guaranteed that at most one in-flight intention can be resident in the shared log at any given point.
In the following discussion, we assume that each component has access to a remote snapshot store (with a key-value or object store API, e.g., S3).
Also, we assume crash failures~\cite{schneider1990smr} for all four components, having discussed the additional Byzantine failure mode of the \Executor{} in the previous sub-section.
We assume that the \agentbus{} is durable and highly available, relying on a distributed shared log~\cite{balakrishnan2012corfu} or implemented as a shim over disaggregated storage.

\textbf{\Decider{}:} The \Decider{} can be viewed as a classical replicated state machine on the shared log.
Its state is a single, compact \Decider{} Policy (the last one written to the log). The \Decider{} can periodically store snapshots of this state to the snapshot store, corresponding to some prefix of the shared log.
On recovery, it simply loads the latest snapshot, and continues processing the log from the corresponding position.
Note that two \Decider{}s can exist at the same time safely: the decision is deterministic, and hence each one will simply append the same decision redundantly to the log.
Downstream components (i.e., the \Executor{}) are expected to ignore duplicate commits for the same intention.

\textbf{\Driver{}:} Likewise, the \Driver{} is also a classical replicated state machine: its state is simply the conversation history, both with the LLM and any external parties (via mail). This state can be maintained locally by the \Driver{} and stored periodically on the remote snapshot store. On recovery, the \Driver{} can restore the snapshot and replay the log. Importantly, since we write inference output entries to the log, replay can be perfectly deterministic despite the non-determinism of the LLM.

Unlike the \Decider{}, however, it is not safe for two \Driver{}s to exist at the same time on the shared log.
Each one will generate redundant intentions in response to external mail or results. Accordingly, when a \Driver{} boots up, its first action is to append a driver policy entry in the shared log electing itself as the designated \Driver{} / fencing any existing \Driver{}. When a \Driver{} observes such an entry on the log from another \Driver{} (and it does not itself have an in-flight election that might land later), it powers itself down.

Note that any components downstream of the \Driver{} (i.e., who are playing intentions) have to also play driver policy entries, so they can reject intentions from a fenced \Driver{}. For example: \Driver{} A appends an intention concurrently with \Driver{} B electing itself; B gets slot 9 while A gets slot 10. Every player of the log has to correctly ignore the intention at slot 10.

\textbf{\Executor{}:} In contrast to the \Decider{} and \Driver{}, the \Executor{} is not a replicated state machine.
Its state can span local state as well as the external environment.
Conventional SMR recovery driven by a snapshot and the shared log is unsafe: commands on the environment are not necessarily idempotent.
Instead, \Executor{} recovery has to be conservative and aim for \textit{at-most-once} execution.
Further, \Executor{}s cannot be relied to drive semantic recovery on their own (e.g., using an LLM to inspect their last intention) without going through \Voter{}s.

Accordingly, when an \Executor{} reboots, it appends a special entry of the result type to the \agentbus{}. This entry is picked up by the \Driver{}; which relays it to the LLM; which responds with new intentions to examine the \bus{}, inspect the environment, and potentially undo/redo actions.
These intentions go through the \Voter{}s to reach the \Executor{}, which actually runs the logic of recovery.
\Executor{}s do not have an explicit election / fence policy entry; the special result acts as an effective fence.
Further, since \Executor{}s interact with the external environment, a new \Executor{} has to either fence the old one away from surrounding services, or alternatively intentions have to rely on idempotent or transactional APIs that can tolerate duplicate invocations.

\textbf{\Voter{}:} As with \Decider{}s and \Driver{}s, \Voter{}s are classical state machines. They store their state in the snapshot store; on recovery, they restore a snapshot and play the log forward. One interesting facet is that our \Decider{} policies refer to types of \Voter{}s rather than individual instances. So, for example, a \Decider{} policy that performs a logical OR on a LLM-based \Voter{} and a rule-based \Voter{} only requires one vote of each type. As a result, \Voter{}s do not need complex fencing logic; they can simply show up and start voting on intents.

\section{\logact{} Implementations}

We now describe different implementations of the two \logact{} layers: the \agentbus{} and the deconstructed agentic state machine above it.

\subsection{\agentbus{} Implementations}

The \agentbus{} API can be implemented on various backends, providing different durability guarantees, performance profiles, and deployment modes.
We implemented three different variants: an in-memory version that does not provide durability; an implementation over SQLite (that stores each entry as a row), providing durability to node reboots but not permanent failures; and a disaggregated variant that stores data on a remote key-value store (with two backing substrates: DynamoDB and an in-house variant called \zippydb{}).

\textbf{\agentbus{} Control Plane.} To simplify the creation and management of \agentbus{} instances, we implemented a control plane component called the \agentkernel{} that runs as a service.
The \agentkernel{} allows clients to create a new \agentbus{} (Raw mode), but also optionally supports spinning up parts of the deconstructed state machine in a remote tier.
For instance, clients creating the \bus{} can optionally (at creation time) start a \Decider{} on it (Auto-\Decider{} mode); run optional \Voter{}s on it from a pluggable library of available \Voter{}s (Auto-\Voter{} mode); or even spawn a sub-agent \Driver{}/\Executor{} using a pre-built image (Spawn mode).
Spawn mode supports multiple backends for sub-agents (K8s, local processes).
As a result, clients have the option of either running the full apparatus of the state machine remotely; or can split the responsibility of running the state machine across the local machine and the remote tier (e.g., running the \Driver{}/\Executor{} and a local \Voter{}, which combines with a remote \Voter{} via some \Decider{} policy).

\subsection{State Machine Implementations}

We explore a number of practical \logact{} implementations (see Table \ref{tab:harnesses}).
Existing harnesses are typically structured as imperative loops; our dirty-slate integration with these harnesses requires us to emulate state machine behavior despite that loop.

\textbf{Dirty-Slate Closed-Source \logact{}: Claude Code}: We modify closed-source Claude Code using pre-execution hooks so that it logs to an \agentbus{} and waits for a voter response.
This is a bare minimum integration for convenience: we are effectively emulating a state machine with an imperative loop.
In this case, we use either Auto-\Decider{} mode or Auto-\Voter{} mode during \agentbus{} creation.

\textbf{Dirty-Slate Open-Source \logact{}: \anonharness{}}: \anonharness{} is an internally used agentic harness that we have source code access to.
Internally, it uses a python interpreter to execute commands (i.e., in CodeAct style).
In this case, we were able to insert \agentbus{} commands surgically to get all entry types on the \bus{}.
However, the code is still an imperative loop rather than a state machine.

\textbf{Clean-Slate \logact{}: \logclaw{}}: We developed our own clean-slate implementation of \logact{} in Rust.
This implementation is a pure state machine (or more accurately, a collection of smaller state machines) on the shared log.
There is no imperative loop; instead, each state machine waits for specific entry types on the log (e.g., the driver for mail; the voter for an intention; etc.), processes each one as it arrives, and appends entry types on the log (e.g., the driver appends intentions; the voter appends votes; etc.).
\logclaw{}'s primary advantage over the other implementations is stronger safety, since it allows the \Driver{} and \Executor{} to be decoupled.

Our harness implementations share some idiosyncrasies common to contemporary LLM-based systems.
Each one has a substantial initial system prompt (e.g., 70KB+ for \anonharness{}).
When we submit inference requests, we re-send the entire inference history (including the initial system prompt plus subsequent deltas) to the inference service.
This is in line with the OpenAI Chat completions API, which is entirely stateless.
We rely on the existence of techniques such as vLLM's Automatic Prefix Caching~\cite{kwon2023vllm} or SGLang's Radix Attention~\cite{zheng2024sglang} to eliminate any redundant inference.
For efficient logging, we only \append{} these deltas to the \agentbus{}, rather than the full inference request each time.

\begin{table}[t]
\centering
\caption{Properties of each \logact{} harness implementation.}
\label{tab:harnesses}
\small
\begin{tabular}{l c c c}
\hline
\textbf{Property} & \textbf{Claude Code} & \textbf{\anonharness{}} & \textbf{\logclaw{}} \\
\hline
Integration & Hooks & Surgical & Native \\
Introspection & Limited & Full & Full \\
\Voter{} Separation & \checkmark & \checkmark & \checkmark \\
\Driver{}/\Executor{} Sep. & & & \checkmark \\
\hline
\end{tabular}
\end{table}

\section{Evaluation}

All experiments run on x86 Linux servers; none of our results depend on low-level hardware performance.
We run agentic harnesses on these servers and use remote commodity inference tiers.
We use a current (as of submission time) frontier model (\frontier{}), as well as an older model from 2024 (\oldermodel{}).
The primary agentic harness used in these experiments is Claude Code; for sub-agents, we use \anonharness{}.
\Voter{}s and \Decider{} are isolated at the process level from the \Executor{}; but the \Driver{} is collocated with it.

\subsection{\logact{} Overhead}

\begin{figure}[t]
\centering
\includegraphics[width=\columnwidth]{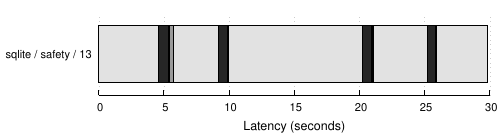}\\[0.25em]
\includegraphics[width=\columnwidth]{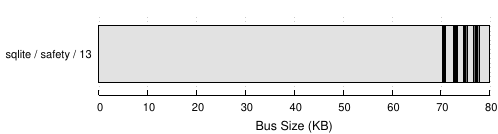}\\[0.25em]
\includegraphics[width=\columnwidth]{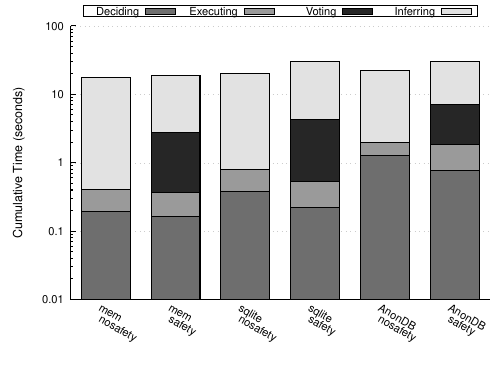}
\caption{\textbf{\logact{} imposes low overhead.} For a simple task (write a C file, compile, run): (Top) most time is spent in inference rather than voting / deciding; (Middle) logging imposes low storage overhead (2.6KB/s); (Bottom) inference continues to dominate even with slower backends.}
\label{fig:exper3}
\end{figure}

We start by characterizing the overhead of logging to the \agentbus{}.
Figure \ref{fig:exper3} shows a \logact{} sub-agent (using \anonharness{} as a harness) performing a simple ``hello world'' task: write a C program, compile it, run it.
As Figure \ref{fig:exper3} (Top) shows, the \logact{} state machine spends most of its time in the \StInferring{} state; with \StVoting{} coming a distant second, and \StDeciding{} not even visible.
\StExecuting{} is also not prominent, though this is a function of the specific task; if we were factoring primes, for example, execution would likely dominate.
Further, Figure \ref{fig:exper3} (Middle) shows that logging every step of the agentic state machine imposes very low strain on the logging substrate: we use up barely 80KB of log in a 30-second agentic task (of which 70KB is the system prompt for \anonharness{}).
Finally, Figure \ref{fig:exper3} (Bottom) has cumulative per-stage latencies for the \logact{} state machine with different backends, with \texttt{on\_by\_default} and \texttt{first\_voter} policies: even with a geo-distributed backend like \zippydb{}, voting and deciding impose little latency overhead relative to inference.

Having established that \logact{} (and the underlying \agentbus{}) imposes low overhead for a typical agentic task, we now explore new capabilities enabled by it.

\begin{figure*}[t]
\centering
\begin{minipage}{0.48\textwidth}
\centering
\includegraphics[width=\linewidth]{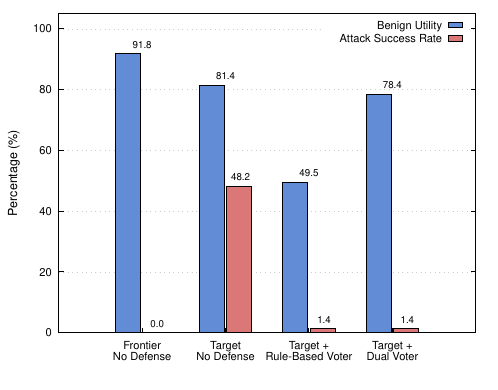}
\end{minipage}
\hfill
\begin{minipage}{0.48\textwidth}
\centering
\includegraphics[width=\linewidth]{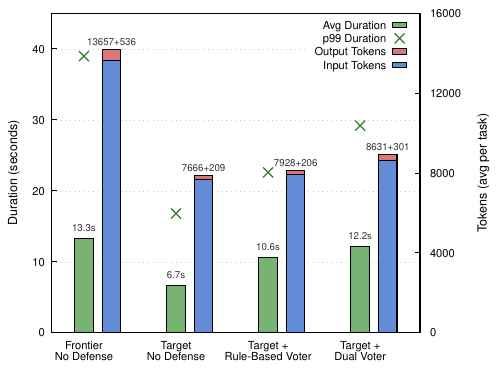}
\end{minipage}
\caption{\textbf{\logact{} stops all unwanted actions with a rule-based \Voter{}; and restores benign Utility with a dual \Voter{} quorum.} Left: \oldermodel{} (No Defense) has high benign Utility, but also an unacceptable ASR; a rule-based \Voter{} stops all unwanted actions (1.4\% involve action-less attacks) but also lowers Utility; while a dual voter quorum (rule-based + LLM) restores Utility. Right: Dual voter adds 82\% latency and uses 13\% more tokens compared to \oldermodel{} (No Defense).}
\label{fig:agentdojo}
\end{figure*}

\begin{figure}[t]
\centering
\includegraphics[width=\columnwidth]{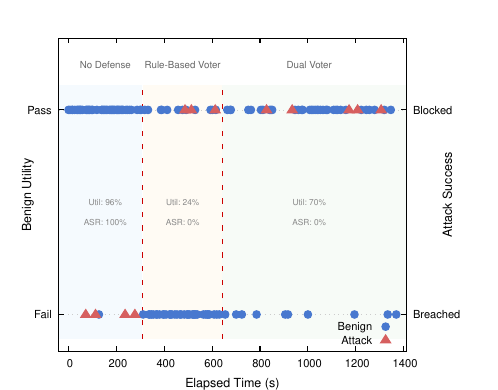}
\caption{\textbf{\Voter{}s can be hot-swapped.} We first add a rule-based \Voter{} to stop attacks; and later, add a second LLM-based \Voter{} to restore Utility. Each switch involves a dynamic change of \Decider{} policy via the \agentbus{}.}
\label{fig:continuous-utility}
\end{figure}

\begin{figure*}[t!]
\begin{minipage}[t]{0.55\textwidth}
\vspace{0pt}
\includegraphics[width=\linewidth]{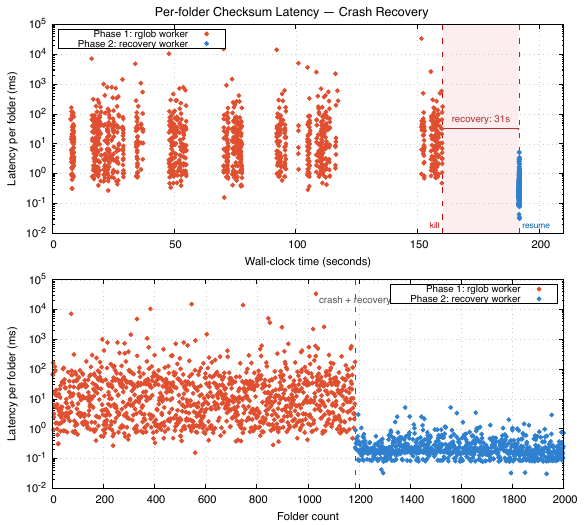}
\end{minipage}%
\hfill
\begin{minipage}[t]{0.43\textwidth}
\vspace{0pt}
\fontsize{6}{7.33}\selectfont
\setlength{\tabcolsep}{2pt}
\begin{tabular}{rrl>{\raggedright\arraybackslash}p{3.6cm}}
\hline
\textbf{\#} & \textbf{Time} & \textbf{Type} & \textbf{Content} \\
\hline
0 & 11:26:35 & mail & Task + crashed agent's \bus{} intentions from orchestrator \\
1 & 11:26:37 & inf-in & Full message history sent to LLM \\
2 & 11:26:42 & inf-out & ``Let me check what was already completed'' \\
3 & 11:26:42 & intent & Code: read output file, count lines \\
4 & 11:26:42 & commit & ON\_BY\_DEFAULT policy (auto-commit) \\
5 & 11:26:42 & result & ``Found 1184 existing lines'' \\
6 & 11:26:42 & inf-in & Updated history sent to LLM \\
7 & 11:26:50 & inf-out & ``Continue from where it left off'' \\
8 & 11:26:50 & intent & Code: list folders, count remaining \\
9 & 11:26:50 & commit & ON\_BY\_DEFAULT \\
10 & 11:26:50 & result & ``1184 done, 4036 total, 816 remaining'' \\
11 & 11:26:50 & inf-in & Updated history sent to LLM \\
12 & 11:26:58 & inf-out & ``Use os.scandir instead of rglob'' \\
13 & 11:26:58 & intent & Code: define + test optimized checksum fn \\
14 & 11:26:58 & commit & ON\_BY\_DEFAULT \\
15 & 11:26:58 & result & ``Test checksum for dsi: 0657473c\ldots'' \\
16 & 11:26:58 & inf-in & Updated history sent to LLM \\
17 & 11:27:07 & inf-out & ``Process all 816 remaining folders'' \\
18 & 11:27:07 & intent & Code: main loop over 816 folders \\
19 & 11:27:07 & commit & ON\_BY\_DEFAULT \\
20 & 11:27:07 & result & ``Processed 816 folders'' (0.36s!) \\
21 & 11:27:08 & inf-in & Updated history sent to LLM \\
22 & 11:27:13 & inf-out & ``Verify the output file'' \\
23 & 11:27:13 & intent & Code: read \& verify final output \\
24 & 11:27:13 & commit & ON\_BY\_DEFAULT \\
25 & 11:27:13 & result & ``2001 lines, DONE in 160.5s'' \\
26 & 11:27:13 & inf-in & Updated history sent to LLM \\
27 & 11:27:21 & inf-out & ``Task completed successfully!'' \\
28 & 11:27:21 & output & Final response \\
\hline
\end{tabular}
\end{minipage}
\caption{\textbf{Agentic introspection enables semantic recovery, health check, and optimization.} Left: Phase~1 (rglob worker) processes $\sim$1200 folders over 155s before being killed. After a 31s recovery window, Phase~2 (recovery worker) processes the remaining $\sim$800 folders. Right: The recovery agent's \agentbus{}. The agent introspects the crashed agent's \bus{}, determines what was completed, diagnoses and fixes the slow implementation, and finishes the remaining 816 folders in 0.36s.}
\label{fig:exper2}
\end{figure*}

\subsection{Pluggable \Voter{}s + Semantic \Voter{}s}

To obtain a representative set of unsafe actions, we used the AgentDojo benchmark~\cite{debenedetti2024agentdojo}.
AgentDojo's setting makes stronger assumptions than ours: user prompts and the baseline LLM are both assumed to be safe; whereas prompt injections via observations on the environments can result in unsafe future actions.
Each test case results in a (Utility, Attack Success) tuple, where Utility is 1 if the agent performs the task requested by the user (and 0 otherwise), while Attack Success is 1 if the agent performs some additional task requested via prompt injection.
We report \textit{benign Utility}: the Utility measured on non-attack test cases; along with the ASR (or Attack Success Rate) for attack test cases.

We first describe our baselines.
\frontier{} performed extremely well on AgentDojo, with 0\% ASR and high Utility (91.8\%); this is to be expected, given that \frontier{} was released after AgentDojo, possibly exposed to it, and likely designed to better handle known attacks.
In this sense, \frontier{} sets the bar for both Utility and ASR (Figure \ref{fig:agentdojo}, Left), as well as task execution latency (Figure \ref{fig:agentdojo}, Right).
In contrast, the older model (\oldermodel{}) was roughly contemporaneous with AgentDojo and unlikely to be trained on it; it provides lower utility at faster speed and lower token usage, but with an unacceptably high ASR of 48.2\%.

Our goal is to demonstrate that \logact{} can add safety to a model that wasn't explicitly trained on AgentDojo's attacks; while providing reasonable Utility, as well as latency / token overhead.
To that end, we evaluated two \logact{} schemes.
In Rule-Based, we use a single \logact{} voter with a large number of rule-based checks.
This is sufficient to drive ASR down to 1.4\% (unsurprisingly, since we created the rules by having an offline LLM look at the \oldermodel{} runs).
The 1.4\% attacks that still succeed are action-less; they involve a simple request/response interaction without any action on the environment, and hence without any intentions for our \Voter{}s to block.
Utility drops considerably compared to \oldermodel{}-No-Defense (49.5\% vs.\ 81.4\%): the static voters unnecessarily block a number of useful actions due to false positives. However, they also allow the agent to successfully complete some actions by preventing the model or environment from getting derailed.

To improve Utility, we deployed a second, LLM-based voter (using the same model \oldermodel{}).
We used a \texttt{boolean\_OR} decider policy: the action commits if either of the voters votes for it.
We specifically prompted the LLM-based voter to act as an override for the static voter: we asked it to examine the \bus{}, including the original user request with the desired task, action outputs containing possible prompt injections, \textit{and also the vote of the rule-based voter}, before making a decision.
With this dual-voter setup, ASR stays at 1.4\%, while benign Utility is improved significantly (78.4\% vs.\ 49.5\% for Rule-Based alone, a 58\% relative improvement).

Figure \ref{fig:agentdojo} (Right) shows the cost of voters.
Rule-based voters add a 58\% slowdown in average latency compared to \oldermodel{} with no defense (10.6s vs.\ 6.7s); while dual-voter adds a further 15\% over rule-based (12.2s vs.\ 10.6s), roughly matching \frontier{} latency (13.3s).
Additional token usage for the dual voter setup is quite low: we only pass in intentions and results to the LLM-based voter, rather than full history; and also only trigger the inference call if the rule-based voter rejects the intent.

The \logact{} design allows new \Voter{}s to be plugged in on-the-fly.
Figure \ref{fig:continuous-utility} shows this capability in action.
We have a single agent processing benign AgentDojo tasks; attacks are injected at a 10\% rate.
Benign utility is plotted on the left y-axis; attack success is plotted on the right y-axis.
In the beginning, we run \oldermodel{} without defenses: utility is high, but all attacks succeed.
At 312s, we change \Decider{} policy to \texttt{first\_voter} and spin up a rule-based \Voter{}.
Attacks are now blocked, but Utility drops as well.
At 655s, we change \Decider{} policy to \texttt{boolean\_OR} and spin up a second, LLM-based \Voter{}.
Utility is restored, while attacks are still blocked.

\subsection{Semantic Recovery / Health Check / Optimization}

To show \logact{}'s ability to recover long-running operations via the \agentbus{}, we use a sample task: we want an agent to scan a large codebase with 2000 top-level folders; and generate a checksum for each top-level folder.
In the experiment, we start an Orchestrator agent (Claude Code), which then creates a Worker sub-agent (\anonharness{}) and sends it a mail asking it to perform the actual task.
We ran multiple iterations of this task (which typically completes in 1-2 minutes) until a particular iteration (the 8th, to be exact) exceeded a 3-minute timeout.
We then started a recovery agent with the following prompt: \prompt{You are recovering from a crash; inspect only the intentions on the original \bus{}; redo the last intention (ideally without repeating work); but fix any obvious reasons that might cause a slowdown in the code.}

It turned out that the slow run used a particular pathological implementation (sorted(rglob(...)) that recursively enumerated every file in the directory tree, and then sorted them before processing; which was particularly slow given that we were using a network-mounted filesystem. The recovery agent realized this via introspection on the \bus{}, and switched to an efficient os.scandir(...) implementation. While we use a separate recovery agent for this experiment, we could equally have restarted the existing worker agent and obtained a similar outcome.

Figure \ref{fig:exper2} (Left) shows latency per folder on the y-axis; and wall-clock time (Left Top) and folder count (Left Bottom) on the x-axis.
The slow agent takes multiple seconds for individual folders, processing 1184 folders before it is killed.
The recovery agent takes 31 seconds to inspect the \bus{} (\textbf{Semantic Recovery:} Entries 1-10), determine the performance bug and suggest an optimization using os.scandir(), which it also locally tests (\textbf{Semantic Health Check:} Entries 11-16), and then continue executing with the faster implementation (Entries 17+), which wraps up the remaining folders in 0.36s (or 290X faster).

\begin{figure}[t]
\centering
\includegraphics[width=\columnwidth]{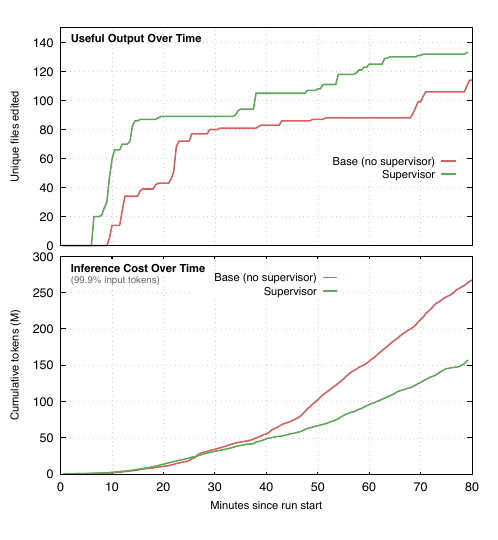}
\caption{\textbf{Agentic introspection makes swarms faster and cheaper.} A supervisor agent inspects each \agentbus{} and sends mail in a 6-agent swarm, resulting in 17\% more work for 41\% less token usage.}
\label{fig:factory}
\end{figure}

\subsection{Optimizing Agent Swarms with Introspection}

To illustrate the benefit of \logact{} and agentic introspection in multi-agent swarms, we deployed a team of one coordinator and six sub-agents (deployed on a K8s cluster) on a large task to add type annotations to a large Python codebase.
We deployed this swarm in two different configurations.
In the Base configuration, each agent uses mailbox messages to coordinate with the others (e.g., to claim pieces of work), but does not use introspection over its \agentbus{}.
In the Supervisor configuration, we deploy an additional agent that periodically inspects the \agentbuses{} of the agents; and sends them mail with useful information.

As Figure \ref{fig:factory} shows, the Supervisor configuration does 17\% more work in a fixed period of time (measured in number of type-fixed files).
Simultaneously, it uses 41\% fewer tokens.
The lower token usage stems from two sources.
At the very beginning of the run, agents struggle with basic infrastructural issues (e.g., building code, finding CLIs, etc.); the Supervisor transmits knowledge of useful fixes from one agent to the other, whereas in Base mode each one redundantly discovers these fixes.
Through the middle of the run, the Supervisor helps agents avoid redundant work (i.e., type-fixing the same files).
In this sense, the Supervisor acts as a centralized ``gossip hub'', ferrying information between agents.
Interestingly, we found a Supervisor to be more effective than a decentralized gossip mode; agents typically did not stick to prompt-driven gossip protocols as their context windows got flooded by task information.

\section{Related Work}

\textbf{Shared Logs.}
The \agentbus{} abstraction builds upon existing shared log APIs~\cite{balakrishnan2012corfu, balakrishnan2013tango, lockerman2018fuzzylog, balakrishnan2020delos, balakrishnan2021delos, jia2021boki, luo2024lazylog, bhat2025fixante, zhu2025impeller} by adding entry types, and providing access control for \append{}/\poll{} operations at the granularity of types.
Prior systems have proposed selective playback based on tags~\cite{balakrishnan2013tango, wei2017vcorfu, jia2021boki}, but mostly as a systems optimization for scaling.
Tango~\cite{balakrishnan2013tango} and Delos~\cite{balakrishnan2021delos} include a notion of state machines sharing a single log; however, these state machines are independent (objects or layers, respectively), rather than deconstructed components of a single state machine.
Types have appeared in publish-subscribe systems~\cite{eugster2007typedpubsub} but not in the shared log or SMR literature.

\textbf{Agentic Safety.}
Camel~\cite{debenedetti2025camel} uses a dual-LLM~\cite{willison2023dualllm} model to separate program generation from program execution, ensuring that only one of two LLMs sees possibly prompt-injected data.
LlamaFirewall~\cite{chennabasappa2025llamafirewall} proposes the use of pluggable safety verifiers; however, it does not describe a full system design for realizing this idea (e.g., with safeguards to ensure that code execution does not rewire the safety framework).
In addition, it stops at safety, whereas \logact{} addresses auditability and recovery as well via its log-based design.
There is also a larger body of work around LLM safety~\cite{inan2023llamaguard} for applications other than agentic systems.

\textbf{Byzantine systems.} Executors in \logact{} can be modeled as Byzantine~\cite{lamport1982byzantine}, in that they execute arbitrary logic generated by an LLM.
However, most practical Byzantine systems focus on the narrower problem of ensuring consensus when a super-majority of replicas are guaranteed to be non-Byzantine.
In our setting of an LLM-driven agent acting on a production environment, it is neither reasonable to make an assumption around how many nodes can act in a Byzantine manner; nor feasible from a cost perspective to execute 3F+1 copies of the agent.
Instead, \logact{} is more closely related to a body of work where untrusted Byzantine logic is audited by trusted verifiers via a trusted log~\cite{haeberlen2007peerreview, chun2007a2m, levin2009trinc}.

\section{Conclusion}

Agentic systems are increasingly powerful and ubiquitous; however, their deployment in real-world production settings is bottlenecked by safety, fault-tolerance, and auditability.
\logact{} structures each agent as a deconstructed state machine playing a shared log: every agentic action is logged before execution, ensuring durability and failure atomicity, as well as subject to configurable safety checks.
Agents can introspect over their execution history on the shared log, enabling semantic recovery from failures, history-aware safety voting, and semantic health checks.
Our evaluation shows that \logact{} adds low overhead compared to inference and execution latency; and provides significant safety and fault-tolerance benefits.
\logact{} is a first step towards ensuring that all agentic activity is visible, stoppable, and recoverable, paving the road for agentic systems that can provide strong guarantees in real-world settings.

\clearpage
\bibliographystyle{ACM-Reference-Format}
\bibliography{bib/references}

\end{document}